\documentclass[jkps,preprint,fleqn,showpacs,showkeys]{revtex4}
\usepackage{graphicx}
\usepackage{amssymb}
\usepackage{amsmath}
\usepackage{bm}
\begin{document}
\setcounter{page}{1}
\title[]{Kondo effect near the Van Hove singularity in biased bilayer graphene}
\author{Stanis{\l}aw \surname{Lipi\'{n}ski}}
\email{lipinski@ifmpan.poznan.pl}
\thanks{Fax: +48-61-868-45-24}
\author{Damian \surname{Krychowski}}
\affiliation{Institute of Molecular Physics, Polish Academy of Sciences\\M. Smoluchowskiego 17,
60-179 Pozna\'{n}, Poland}


\begin{abstract}
Magnetic impurity adsorbed on one of the carbon planes of a bilayer graphene is studied. The formation of the many-body $SU(2)$ and $SU(4)$ resonances close to the bandgap is analyzed within  the mean field Kotliar-Ruckenstein slave boson approach. Impact of enhanced hybridization and magnetic instability of bilayer  doped near the Van Hove singularity on the screening of magnetic moment is discussed.
\end{abstract}



\maketitle

\section{INTRODUCTION}
Bilayer graphene (BLG) is a quasi-two-dimensional gapless chiral electron-hole system with parabolic bands with conduction and valence branches touching at a point \cite {Sarma11}. The AB Bernal stacked bilayer has inversion symmetry. The non-zero bandgap can be induced by breaking this symmetry e.g. by applying transverse electric field. As opposed to conventional semiconductors, bandgap of BLG can be tuned, what is the reason of  technological importance of this material \cite {Zhang05}. The density of states (DOS) of biased BLG has the square-root  singularities at the band edges, just like in one-dimensional systems. It is well known that  electronic instabilities are expected at the crossing of the Fermi level ($E_{F}$) with Van Hove singularity (VHS). The possibility of magnetic ordering  in bilayer graphene   has been discussed in several papers  \cite {Nilsson06,Stauber07,Castro08}. Since the chemical potential of BLG can be tuned one expects that  formation of local magnetic moment can be also controlled by gate voltage. A strong enhancement of DOS at the Fermi level increases hybridization, what moves impurity state  towards mixed valence (MV) regime or even suppresses the many body resonance. It is a challenging problem to discuss  Kondo or mixed valence resonance under above mentioned conditions taking also into account  possible magnetic instability in BLG matrix.
Understanding of the interplay of these competitive  effects is important for prospect of  Van Hove singularity engineering.

\section{Model}
Bilayer graphene \cite {Sarma11} is composed of two coupled honeycomb lattices of carbon atoms. We consider  AB-Bernal stacking, where the top layer has its A sublattice on top of sublattice B of the bottom layer. Indices 1 and 2 are used to label top and bottom layers, respectively. In the tight binding approximation the BLG Hamiltonian reads:
\begin{eqnarray}
{\cal H}_{BLG}=&&\sum_{p\alpha i \sigma} -(-1)^{\alpha}(V_{g}/2) n^{(p)}_{\alpha i\sigma}+{\cal{U}}_{g}\sum_{p\alpha i \sigma}n^{(p)}_{\alpha i\uparrow}n^{(p)}_{\alpha i\downarrow}-t\sum_{\langle ij\rangle\alpha}(a^{+}_{\alpha i\sigma}b_{\alpha j\sigma}+h.c.)
\nonumber\\&&-t_{\perp} \sum_{i}(a^{+}_{1i\sigma}b_{2i\sigma}+h.c.),
\end{eqnarray}
where $a_{\alpha i\sigma}$($b_{\alpha i\sigma}$) annihilates electron on sublattice $A$($B$) in plane $\alpha =1,2$ at site $i$, $\langle ij\rangle$ represent a pair of in-plane nearest neighbors, $p$ numerates sublattices, $t$, $t_{\perp}$ denote  the in-plane and interlayer  hopping integrals respectively, ${\cal{U}}_{g}$  measures  on-site graphene   Coulomb interaction and $V_{g}$ is the  electrostatic bias applied to the two planes. The value of $t$ ($t = 2.7$ eV) is taken as the energy unit and we assume $t_{\perp} = 0.2$, ${\cal{U}}_{g} =1$ \cite {Castro08}. It is out of scope of the present paper to discuss the details of magnetic instabilities near VHS (see for e.g. \cite {Nilsson06,Stauber07,Castro08}). For simplicity, following \cite {Castro08} we adopt the  mean field approach and define ferromagnetic  ground state characterized by unequal magnetizations of the layers and different electron densities:
\begin{eqnarray}
\langle n^{(p)}_{\alpha i \sigma}\rangle=\frac{n-(-1)^{\alpha}\Delta n}{8}+\sigma\frac{m-(-1)^{\alpha}\Delta m}{8},
\end{eqnarray}
where $n$ and $m$ denote electron density and magnetization per unit cell respectively and $\Delta n$, $\Delta m$ the corresponding differences between the layers.
The analysis presented in this  paper is addressed to TM adatoms located in the energetically favorable position -centre of a hexagon (hollow site).  The symmetry of hollow site in BLG is $C_{3v}$ and five d orbitals split into two orbital doublets corresponding to two-dimensional  representation ($E$) and a singlet of one-dimesional representation $A_{1}$. In  Kondo physics  the crucial role play electrons with unpaired spins. To get a microscopic insight  into the orbital occupations  of the concrete impurity  ab initio calculations are required \cite {Wehling10}.  In our model analysis based on Anderson Hamiltonian the picture is simplified by considering only the pair  of orbitally degenerate states  ($SU(4)$) or  by discussing even simpler case when symmetry is further reduced to two spin-orbitals  ($SU(2)$) \cite {Wehling10}.
The impurity Hamiltonian for $SU(4)$ symmetry reads:
\begin{eqnarray}
{\cal H}^{|m|}_{imp}=\sum_{m\sigma}\epsilon_{0}n_{m\sigma}+{\cal{U}}_{0}\sum_{m}n_{m\uparrow}n_{m\downarrow}
+V^{|m|}\sum_{m\sigma}(c^{+}_{m\sigma}d_{m\sigma}+h.c.),
\end{eqnarray}
with $|m|=1,2$, $\epsilon_{0}=-0.5$. $c_{m\sigma}=(1/\sqrt{6})\sum_{i}(e^{\imath m\varphi_{iA}}a_{1i\sigma}+e^{\imath m\varphi_{iB}}b_{1i\sigma})$ and $\varphi_{iA(B)}$ denotes the angle between a fixed crystalline axes and the bond from site $i$ to the impurity.
To discuss the many-body problem we use mean field Kotliar-Ruckenstein slave boson approach \cite {Kotliar86} in the infinite impurity Coulomb interaction limit  ${\cal{U}}_{0}\rightarrow\infty$. The finite ${\cal{U}}_{0}$ limit will be discussed elsewhere.  Three introduced auxiliary boson fields $e$, $p_{m\sigma}$ ($\sigma=\pm1$)  project respectively onto the empty or single occupied state (with spin up or down). The slave boson Hamiltonian reads:
\begin{eqnarray}
{\cal H}_{imp}=&&\sum_{m\sigma}\epsilon_{0}f^{+}_{m\sigma}f_{m\sigma}+
V^{|m|}\sum_{m\sigma}(c^{+}_{m\sigma}z_{m\sigma}f_{m\sigma}+h.c.)
\nonumber\\&&+\lambda(I-1)+\sum_{m\sigma}\lambda_{m\sigma}(Q_{m\sigma}-f^{+}_{m\sigma}f_{m\sigma})
\end{eqnarray}
The terms with Langrange multipliers $\lambda$ and $\lambda_{m\sigma}$ are incorporated to prevent double occupancy and to fulfill the complete relations.  $I=e^{+}e+\sum_{m\sigma}p^{+}_{m\sigma}p_{m\sigma}$, $Q_{m\sigma}=p^{+}_{m\sigma}p_{m\sigma}$, $z_{m\sigma}=e^{+}p_{m\sigma}/(\sqrt{Q_{m\sigma}}\sqrt{1-Q_{m\sigma}})$ and pseudofermion operator $f_{m\sigma}=z_{m\sigma}d_{m\sigma}$. The slave boson parameters $e$, $p_{\sigma}$ ($p_{\sigma}=p_{m\sigma}=p_{-m\sigma}$), $\lambda$ and $\lambda_{\sigma}$ are determined minimizing the ground state energy of (4).

\section{Results and discussion}
\begin{figure}
\includegraphics[width=8 cm,bb=30 120 550 660,clip]{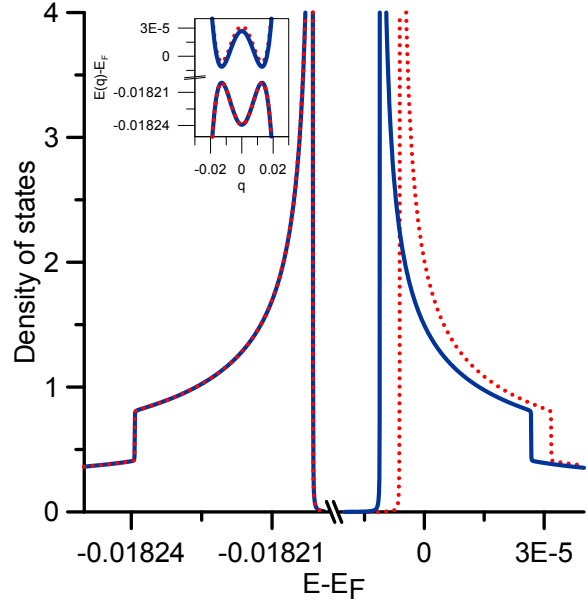}
\caption{(Color online) Spin-resolved densities of states of BLG close to the bandgap plotted by solid (dotted) line for up (down) spin. Inset shows the corresponding Hartree-Fock bands, $V_{g}=0.0185$.}\label{fig1}
\end{figure}

\begin{figure}
\includegraphics[width=8 cm,bb=30 120 550 660,clip]{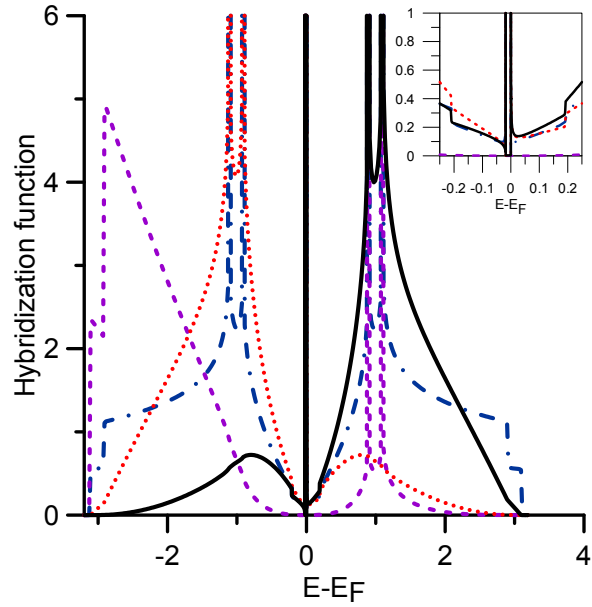}
\caption{(Color online) Rescaled  imaginary parts of hybridization functions for the impurity in the hollow site for three angular momentum symmetries $\Sigma_{m}/V^{2}$ ($m =0$ - dashed line, $m =1$ - dotted line, $m =2$ - solid line) compared with BLG density of states (dash dash dotted line). Inset shows the same hybridization functions as in the main picture, but in the narrower energy range. The assignment of the lines is preserved.}\label{fig2}
\end{figure}

Fig. \ref{fig1} illustrates DOS of the mean field  ferromagnetic state of BLG ($\varrho_{\sigma}(E)$) for Fermi level located close to the upper gap edge. The spin splitting of near-gap conduction band states is visible. The low energy  band structure  presented in the inset shows a \emph{Mexican hat} dispersion.  The wave vector $q$  measures the deviation  from the centre of the valley $K_{\pm}=\pm(4\pi/3a,0)$, $(k=K_{\pm}+q)$.  For the assumed positive potential applied to plane $1$ the wavefunction  amplitudes  of these states are  higher for layer $1$ than for layer $2$.  Even though  the induced magnetic moments are extremely small, the spin polarization at the Fermi level  $P_{BLG}=\frac{\varrho_{\uparrow}(E_{F})-\varrho_{\downarrow}(E_{F})}{\varrho_{\uparrow}(E_{F})
+\varrho_{\downarrow}(E_{F})}$ might be strong if $E_{F}$ is placed close to the upper edge. In the  narrow energy range in this region  the  full spin polarization of electrons is observed.  Apart from Van Hove singularities in the position of minima (maxima) of conduction band (valence band) also jumps of DOS are observed.  They correspond to maximum (minimum) of energy in the centers of valleys (see the inset).
Fig. \ref{fig2} shows hybridization self energies for angular momentum symmetries  $m = 0,1,2$. The presented hybridization function $\Sigma_{m}$ plays the role of self-energy of the impurity Green's function $G_{0}(\omega)$ corresponding to (3), $\Sigma_{m}(\omega)=\omega-\epsilon_{0}-G^{-1}_{0}(\omega)$. As it is seen for hollow site  hybridization of $m = 0$ symmetry  is strongly suppressed  in the low energy range.
The self-energies for $m = 1$ and $m = 2$ do not differ very much in this region, but for higher energies distinctive differences  are observed and the dependences are strongly asymmetric with respect to the zero energy.  Very roughly the electron part of $m = 1$ self energy corresponds to the hole part of $m = 2$.  The observed splitting of the peaks around $E-E_{F}\approx1.03$ corresponding to $M$ point of Brillouin zone are caused by interlayer coupling.
\begin{figure}
\includegraphics[width=8 cm,bb=30 120 550 660,clip]{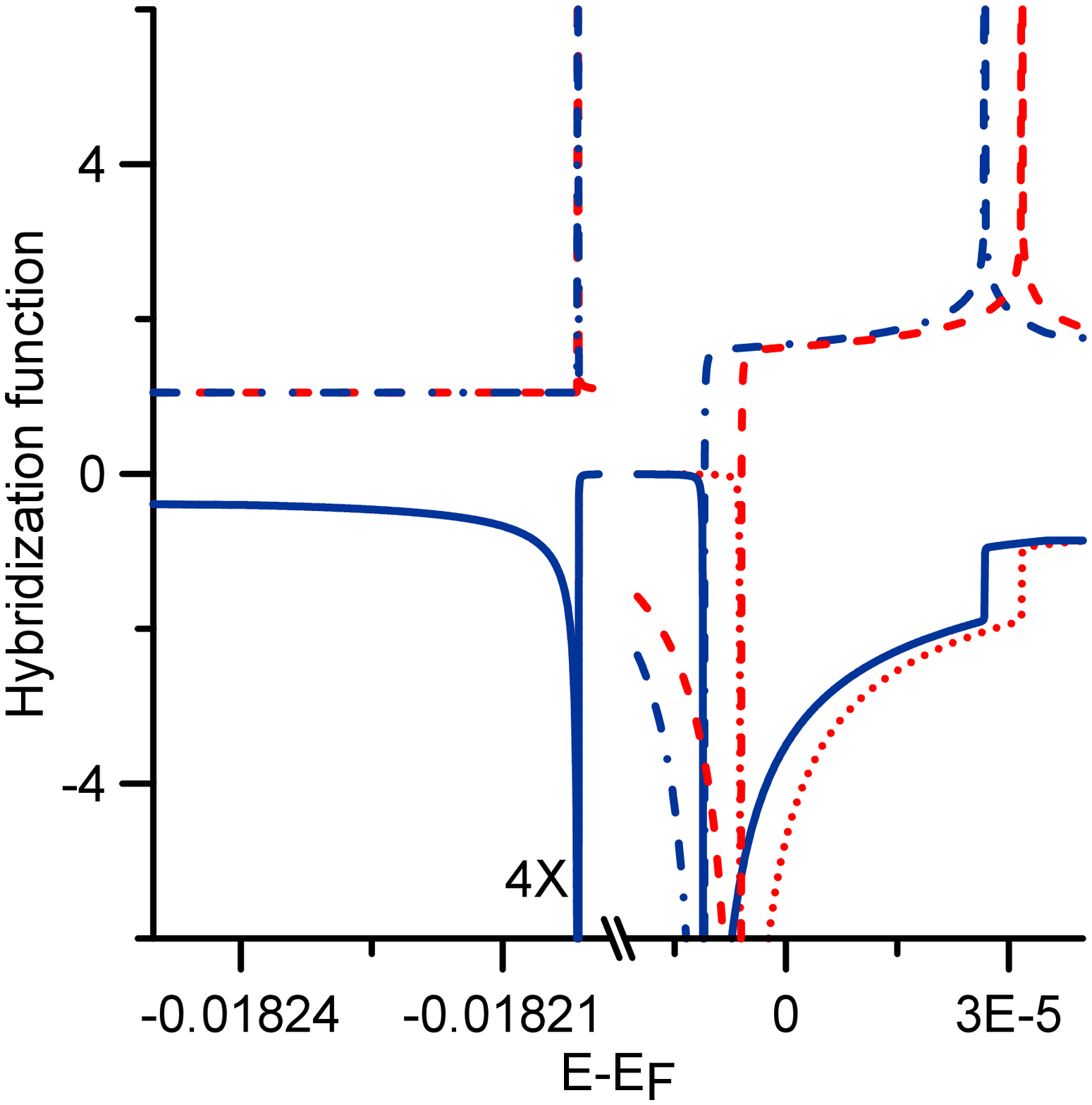}
\caption{(Color online) Real and imaginary parts of rescaled hybridization functions $\Sigma_{1\sigma}/V^{2}$
plotted   close to the bandgap. Solid and dotted lines represent imaginary parts for spin up and spin down respectively and  dash dash dotted and dashed lines represent the real parts of the spin up and spin down self energies. For transparency imaginary part of hybridization function below the lower bandgap edge is extended four times.}\label{fig3}
\end{figure}
The details of the spin dependent  structure around band gap are presented on Fig. \ref{fig3}, where in  addition to imaginary parts also   real parts of self energies are plotted. Divergence of imaginary part for $E_{F}$ located at the edge signals non-Fermi liquid  behavior.
\begin{figure}
\includegraphics[width=8 cm,bb=0 0 720 720,clip]{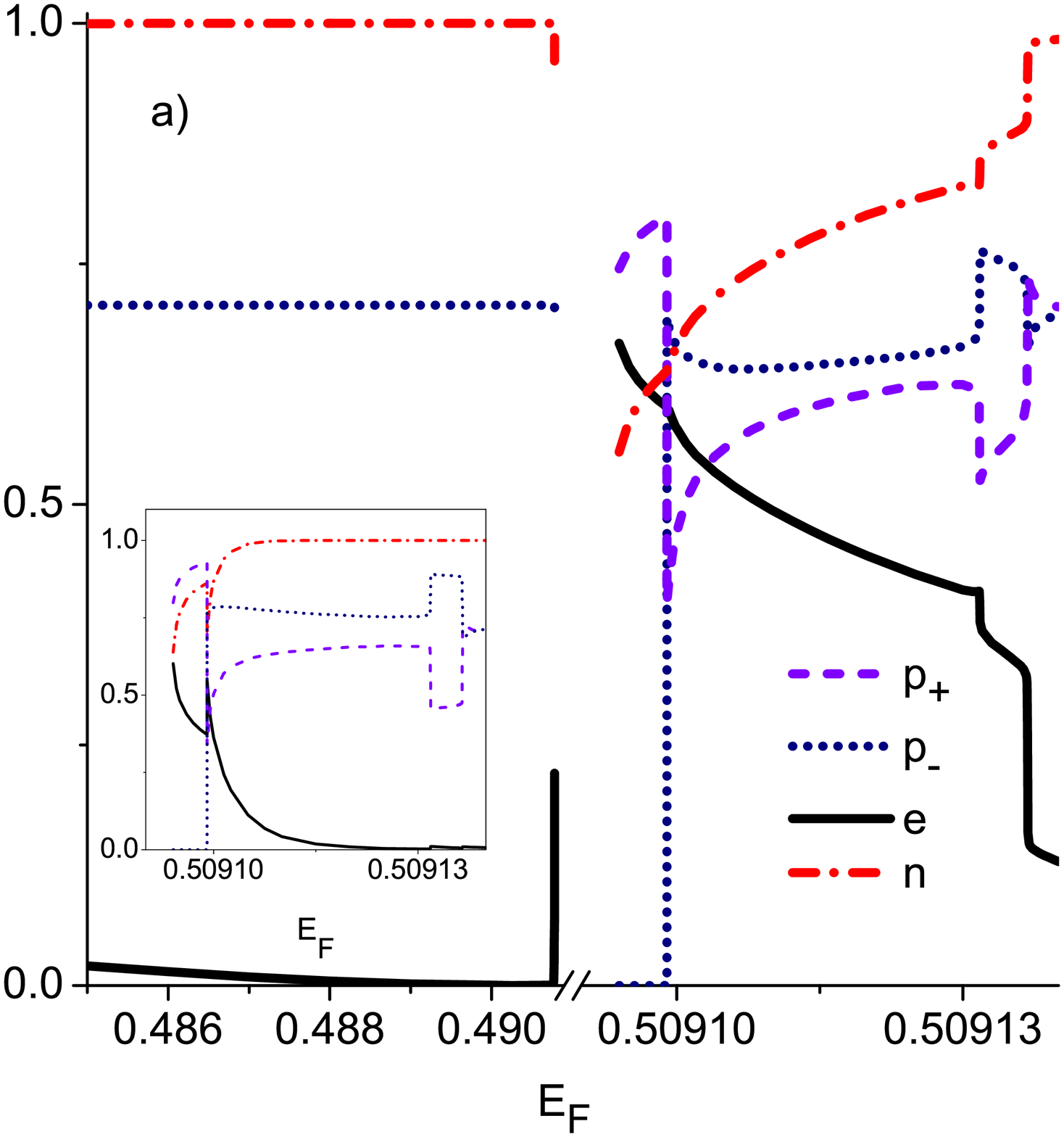}
\includegraphics[width=8 cm,bb=0 0 720 720,clip]{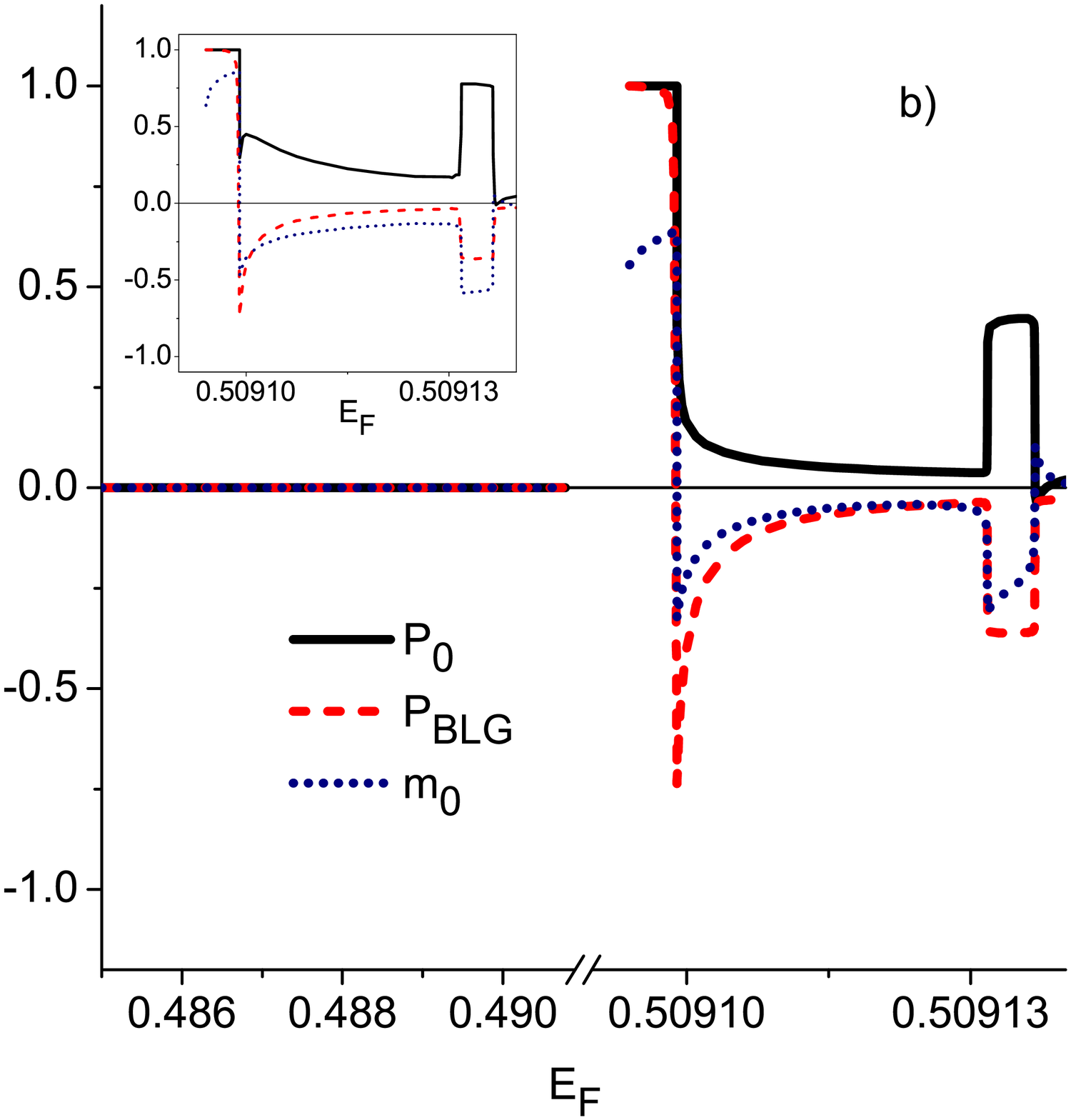}
\caption{(Color online) Evolution of  $SU(2)$ impurity state with the shift of the Fermi level in the narrow energy range around bandgap   a) The expectation values of slave-boson operators $e$, $p_{\sigma}$ and the impurity electron occupation $n$ for coupling parameter  $V = 0.25$.  b) Polarization of impurity $P_{0}$ , impurity magnetization $m_{0}$ and  polarization  of BLG  ($P_{BLG}$). Insets of both pictures present the corresponding dependences for $V = 0.1$ for $E_{F}$ located above the upper gap edge.}\label{fig4}
\end{figure}
The observed  peaks and jumps of the  real part substantially renormalize   the effective impurity energy level in this region.  We do not  address here  the subtle  problem of local moment formation in BLG, this has been recently discussed e.g. in \cite {Killi11}.  We have only checked that in Hartree-Fock approximation the magnetic moment of impurities do not vanish for the assumed parameters.
\begin{figure}
\includegraphics[width=8 cm,bb=0 0 720 720,clip]{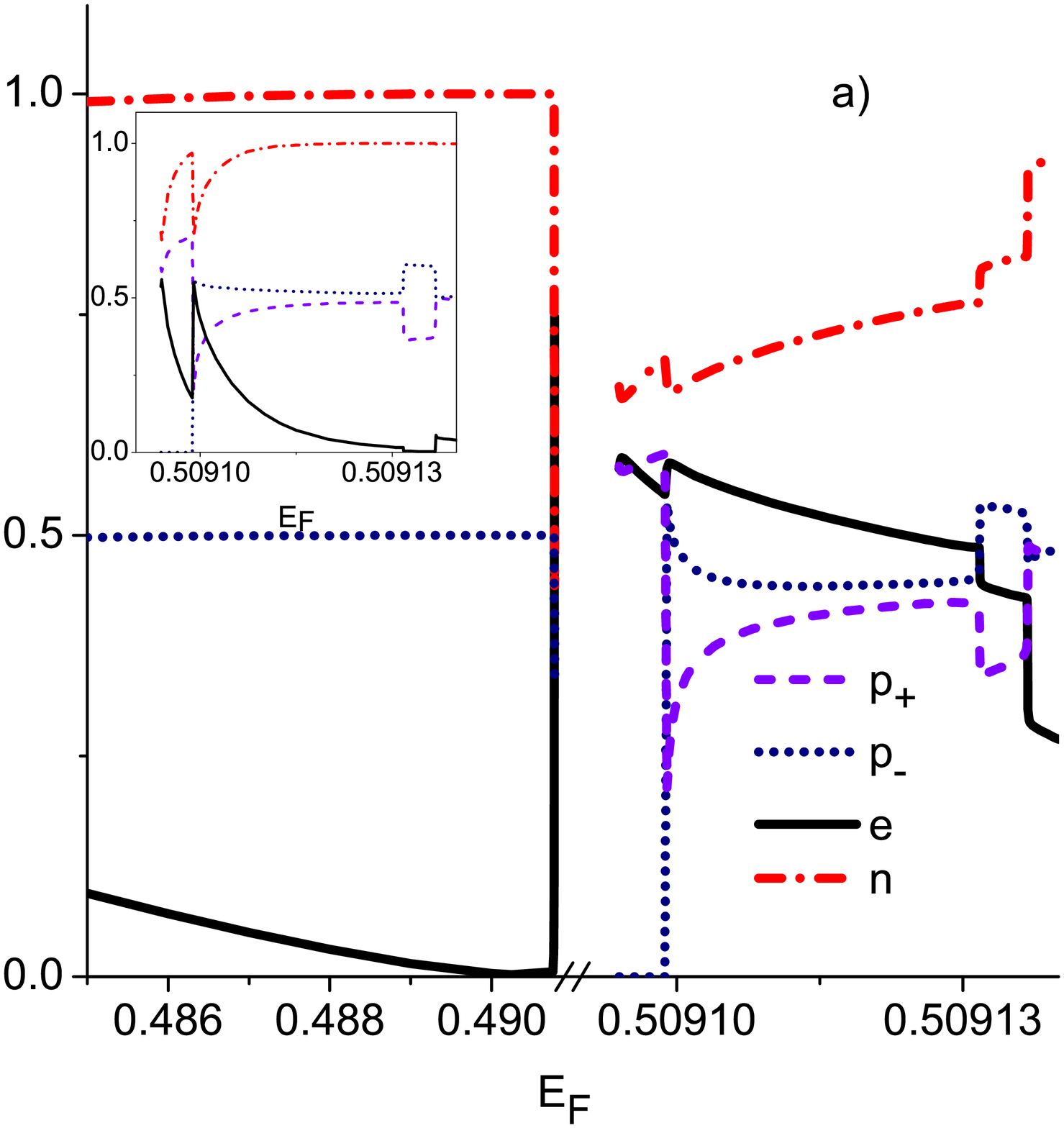}
\includegraphics[width=8 cm,bb=0 0 720 720,clip]{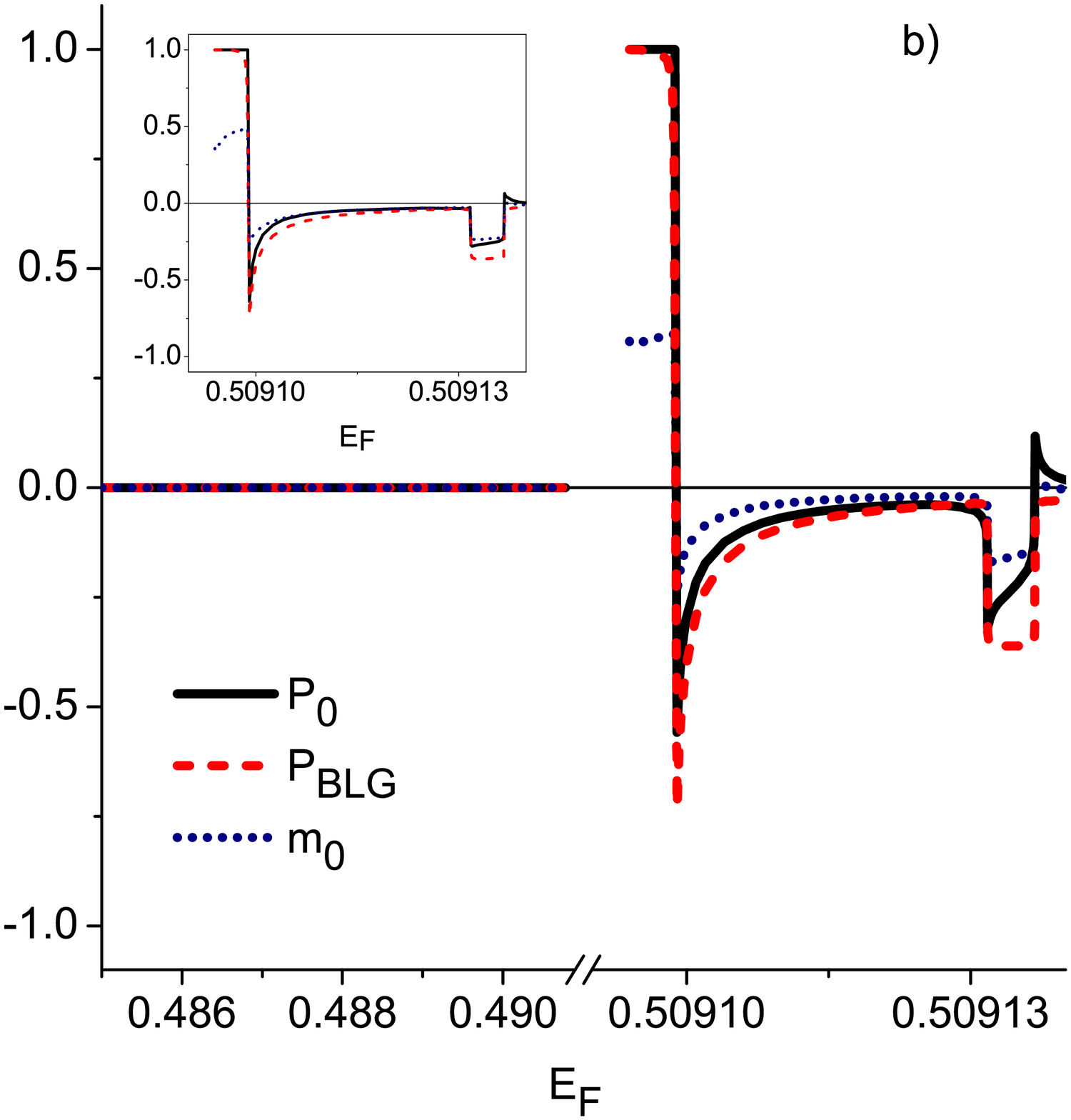}
\caption{(Color online) The same as in Fig. 4 but for SU(4) symmetry.}\label{fig5}
\end{figure}
Figures \ref{fig4} and \ref{fig5} show evolution of the expectation values of slave boson operators, magnetic moment and polarization of impurity versus $E_{F}$ for $SU(2)$ and $SU(4)$ symmetries. The factors determining  the character of the many-body resonance are the deepness of the atomic level with respect to $E_{F}$ and hybridization strength.
\begin{figure}
\includegraphics[width=10 cm,bb=30 60 550 700,clip]{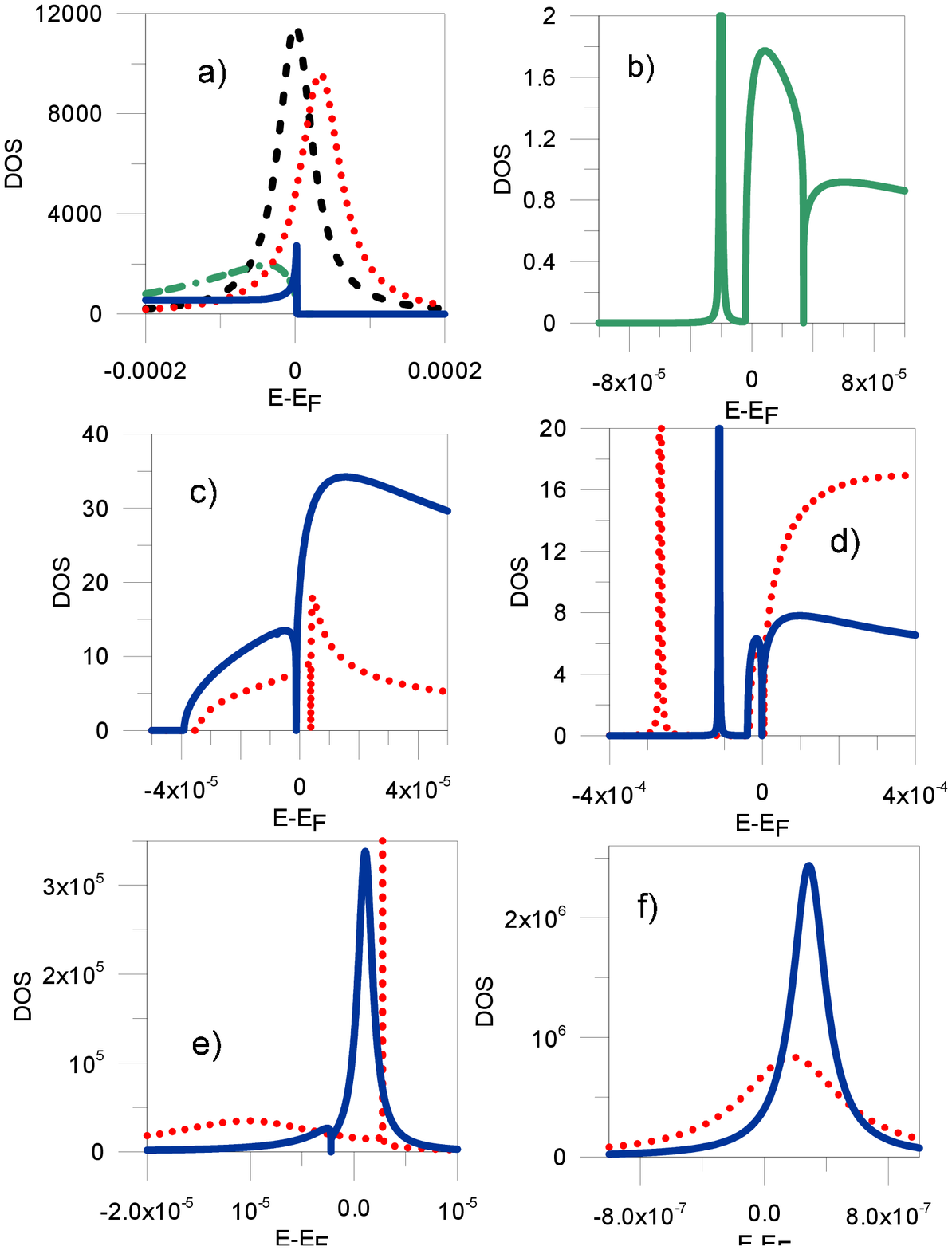}
\caption{(Color online) Representative impurity densities of states a)  $V = 0.25$,  $E_{F}$ located below lower bandgap edge (unpolarized BLG): $SU(2)$  $E_{F}\approx0.48580$ dashed line, $SU(2)$ $E_{F}\approx0.49080$ dash dash dotted line, $SU(4)$  $E_{F}\approx0.48580$ dotted line, $SU(4)$  $E_{F}\approx0.49080$  solid line, b) $V = 0.25$, DOS of orbital MV state $SU(4)\rightarrow SU(2)$ for $E_{F}\approx0.50909$ (fully spin polarized BLG), c)  $V = 0.25$, $SU(2)$, $E_{F}$ located  between the  upper bandgap edge and nearby step in BLG DOS (polarized BLG) ($E_{F}\approx0.50913$), solid line for spin up, dotted line for spin down, d) the same as (c), but for $SU(4)$ symmetry, e) the same as (c), but for $V=0.1$, f) the same as (d), but for $V=0.1$.}\label{fig6}
\end{figure}
For the narrow energy range discussed, where the dramatic changes of the latter are observed,  impact of the hybridization  is dominant. The representative densities  of states of the impurity ($\varrho_{0\sigma}(E)$) are shown on Figures \ref{fig6}. Below the lower gap edge, where hybridization self-energy is small, structureless and spin independent (Figs. \ref{fig4}a, \ref{fig5}a), typical Kondo resonance is formed centered at $E_{F}$ for $SU(2)$ symmetry and shifted above $E_{F}$ for $SU(4)$ case (Fig \ref{fig6}a).
When $E_{F}$ moves closer to the edge and both real and imaginary parts of self-energy are strongly enhanced broadening of many-body spectral function results for both symmetries and for $SU(4)$ additional structure at the edge is  also visible reflecting a new pole in the  impurity Green's function occurring close to the edge, but still in the region of finite density of states of BLG. Within the gap  a resonance can be also formed, and in this case the resonance is not masked by finite DOS and takes the form of Dirac delta like line (Figs. \ref{fig6}b,d). Of interest are also dips in the  impurity DOS  (Figures \ref{fig6}b-e), occurring for energies where steps in the imaginary parts and peaks in real parts of self energies are observed  (Fig. \ref{fig3}).
The narrow energy region above the upper edge is of special interest for spintronics, because spin polarization of BLG matrix results in  spin splitting of resonances (Figs. \ref{fig6}c-f). For typical  coupling parameter corresponding to the equilibrium distance of  impurity from the graphene layer (assumed value $V = 0.25$ corresponds to Co impurity-graphene coupling \cite {Wehling10}), the calculated average occupation considerably deviates from $n = 1$ i.e. impurity is in mixed valence state for both symmetries (Figs. \ref{fig6}c,d, Figs. \ref{fig4}a,\ref{fig5}a).
More interesting is the case of a weaker coupling, when the  spin polarized Kondo state is formed  (Figs. \ref{fig6}e,f). Possible reduction of the coupling can be achieved e.g. by chemical engineering (ligands), or introducing ultrathin dielectric film separating impurity from graphene layer, or putting the adatom further from the surface by STM methods  \cite {Ternes09}. For Fermi level located between  the upper edge and  nearest  higher in energy  step in BLG density of states the  impurity polarization for SU(2) symmetry is opposite to BLG polarization, whereas  for  $SU(4)$  has the same sign (Figs. \ref{fig4}b,\ref{fig5}b).  This tendency is observed both in MV and in Kondo states and is in agreement with earlier calculations for regular density of states of the electrodes performed with the use of different techniques \cite {Martinek03,Choi04,Lipinski11}. This behavior  reflects the fact that despite  the same sign of the spin  splitting for both symmetries (opposite to BLG spin splitting), the dominance of spin resolved densities of states at the Fermi level is different  due to significantly different location of unperturbed  many-body peaks  in $SU(2)$ and $SU(4)$ cases.
Interesting observation is  a possibility of full spin-polarization of  impurity state in the region of full polarization of BLG matrix and reverse of magnetic moment when  $E_{F}$ enters the region of full BLG spin polarization. Since the  Fermi level can be shifted  by gate voltage this gives the way of electric control of magnetic moment, what is a highly desirable property for spintronics.

In summary, present investigation only signals the interplay of different factors on the many-body physics of magnetic impurity in bilayer graphene doped near the Van Hove singularity at the bandgap edges.  The exact description of  the crossing of the Fermi energy with a Van Hove singularity in the DOS  would require more sophisticated methods, both in defining the magnetic state of the  bilayer going beyond mean field scheme and in discussion of screening processes, where  analysis outside  SBMFA is required. In addition to hole-like processes discussed above also electron-like charge fluctuations might play the role in reconstruction of Kondo resonance and this topic will be the subject of a forthcoming paper.

\begin{acknowledgments}
This work was supported by the Polish Ministry of Science and Higher Eduction as a research project No. N N$202199239$ in years $2010-2013$.
\end{acknowledgments}


\begin{references}
\bibitem{Sarma11} S. Das Sarma, S. Adam, E.H. Hwang, and E. Rossi, Rev. Mod. Physics \textbf{83}, 407 (2011).
\bibitem{Zhang05} Y. Zhang, T.-T. Tang, C. Girit, Z. Hao, M. C. Martin, A. Zettl, M. F. Crommie, Y. R. Shen and F. Wang, Nature \textbf{459}, 820 (2009).
\bibitem{Nilsson06} J. Nilsson, A. H. Castro Neto, N. M. R. Peres, and F. Guinea, Phys. Rev. B \textbf{73}, 214418 (2006).
\bibitem{Stauber07} T. Stauber, N. M. R. Peres, F. Guinea, and A. H. Castro Neto, Phys. Rev. B \textbf{75}, 115425 (2007).
\bibitem{Castro08} E. V. Castro, N. M. R. Peres, T. Stauber, and N. A. P. Silva, Phys. Rev. Lett. \textbf{100}, 186803 (2008).
\bibitem{Wehling10} T. O. Wehling, A. V. Balatsky, M. I. Katsnelson, A. I. Lichtenstein, and A. Rosch, Phys. Rev. B \textbf{81}, 115247 (2010).
\bibitem{Kotliar86} G. Kotliar and A. E. Ruckenstein, Phys. Rev. Lett. \textbf{57}, 1362 (1986).
\bibitem{Killi11} M. Killi, D. Heidarian, and A. Paramekanti, New J. Phys. \textbf{13}, 053043 (2011).
\bibitem{Ternes09} M. Ternes, A. J. Heinrich, and W. D. Schneider, J. Phys.: Condens. Matter \textbf{21}, 053001 (2009).
\bibitem{Martinek03} J. Martinek, Y. Utsumi, H. Imamura, J. Barna\'{s}, S. Maekawa, J. K\"{o}nig, and G. Sch\"{o}n, Phys. Rev. Lett. \textbf{91}, 247202 (2003).
\bibitem{Choi04} M. S. Choi, D. Sanchez, and R. L\'{o}pez, Phys. Rev. Lett \textbf{92}, 056601 (2004).
\bibitem{Lipinski11} S. Lipi\'{n}ski and D. Krychowski, Phys. Rev. B \textbf{81}, 115327 (2011).
    \end{references}
\end{document}